\documentclass[12pt]{article}
\usepackage{amsmath}
\usepackage{amssymb}
\usepackage{slashed}
\usepackage{graphicx}
\usepackage{subfigure}
\usepackage{cite}
\usepackage{url}
\usepackage[small]{caption}

\begin{document}
\baselineskip 0.6cm

\def\bra#1{\langle #1 |}
\def\ket#1{| #1 \rangle}
\def\inner#1#2{\langle #1 | #2 \rangle}

\begin{titlepage}

\begin{flushright}
UCB-PTH-15/13\\
\end{flushright}

\vskip 1.5cm

\begin{center}
{\Large \bf Axion Isocurvature and Magnetic Monopoles}

\vskip 0.7cm

{\large Yasunori Nomura, Surjeet Rajendran, Fabio Sanches}

\vskip 0.4cm

{\it Berkeley Center for Theoretical Physics, Department of Physics,\\
 University of California, Berkeley, CA 94720, USA}

\vskip 0.1cm

{\it Theoretical Physics Group, Lawrence Berkeley National Laboratory,
 CA 94720, USA}

\vskip 0.8cm

\abstract{We propose a simple mechanism to suppress axion isocurvature 
 fluctuations using hidden sector magnetic monopoles.  This allows for 
 the Peccei-Quinn scale to be of order the unification scale consistently 
 with high scale inflation.}

\end{center}
\end{titlepage}

\section{Introduction}
\label{sec:intro}

Cosmic inflation not only provides a framework to address many puzzles 
of early universe cosmology~\cite{Guth:1980zm,Linde:1981mu} but also 
incorporates a mechanism that seeds the formation of the structure 
in the universe~\cite{Hawking:1982cz}.  An exciting aspect of the 
inflationary mechanism is that it also sources gravitational 
waves.  If inflation occurs at a sufficiently high scale ($\sim 
10^{15}\mbox{--}10^{16}$ GeV), the amplitude of these gravitational 
waves is large enough to leave a measurable imprint on the polarization 
of the cosmic microwave background (CMB)~\cite{Zaldarriaga:1998ar}. 
A number of CMB polarization experiments are presently searching for 
this signal~\cite{Ade:2014xna}.  A positive signal in such an experiment 
would have interesting implications for particle physics, especially 
for ultra-light bosonic fields.  Bosonic fields with masses lighter 
than the inflationary Hubble scale are efficiently produced by inflation 
and can cause isocurvature perturbations in the CMB~\cite{Linde:1984ti}. 
High scale inflation thus leads to interesting constraints on ultra-light 
bosons, including the QCD axion provided the axion decay constant 
$f_a$ is greater than the inflationary scale.

It is widely regarded~\cite{Fox:2004kb} that a discovery of inflationary 
gravitational waves would rule out the QCD axion with a decay constant 
$f_a \gtrsim 10^{16}~{\rm GeV}$, a range that is favored by several 
theoretical considerations~\cite{Svrcek:2006yi}.  Experiments have also 
been proposed recently to search for the QCD axion in this parameter 
range~\cite{Graham:2011qk}, and it is of great interest to delineate 
the viable parameter space accessible to these efforts.  For example, 
this bound disappears if the QCD axion acquires a large mass during 
inflation, damping the production of isocurvature modes.  At the end 
of inflation, however, this mass has to nearly vanish for the QCD axion 
to solve the strong $CP$ problem.  While models achieving this do exist 
(see~\cite{Dvali:1995ce} for example), they face the difficulty that 
the mechanism responsible for generating a large axion mass during 
inflation has to violate the Peccei-Quinn symmetry while ensuring that 
this violation remains sufficiently sequestered from the axion after 
inflation.  This task is made even more difficult by the fact that 
these dynamics must couple to the inflaton.  Other proposals to alleviate 
the tension between high scale inflation and the QCD axion include 
a dynamically changing Peccei-Quinn breaking scale~\cite{Kaplan:2005wd}. 
While reasonable, such models sacrifice some of the theoretical 
arguments underlying high $f_a$ axions.  There are also attempts 
that involve transfer of the axion isocurvature from one species 
to another~\cite{Kitajima:2014xla}, but these typically deplete 
the dark matter abundance of the axion, eliminating one of the 
promising ways to search for them.  It might also be possible to 
relax these constraints by dumping entropy into the universe around 
the QCD phase transition~\cite{Kawasaki:2014una}, but these channels 
are constrained~\cite{Fox:2004kb}.

In this paper, we investigate an alternative possibility:\ what if the 
QCD axion acquires a large mass {\it after} inflation, which subsequently 
disappears {\it before} the QCD phase transition?  If this mass is larger 
than the Hubble scale during a large interval, somewhere between the 
reheating and QCD scales, then the axion field oscillates earlier and 
the fluctuations in the field will be damped, relaxing into the minimum 
of the potential generating this large mass.  When this mass (and 
potential) subsequently disappears, the average axion field takes 
a value corresponding to this minimum.  Since this minimum is in 
general displaced from the QCD minimum, the misalignment between 
these two points regenerate a cosmic abundance of the QCD axion when 
the axion reacquires a mass during the QCD phase transition, enabling 
it to be dark matter.  The isocurvature perturbations, however, will 
be small since the initial evolution of the field causes the perturbations 
to coalesce around the initial minimum, while the subsequent dark 
matter abundance is generated by the homogeneous misalignment between 
the QCD minimum and the initial minimum.

How can we give such a large initial mass that then disappears almost 
completely?  We accomplish this by coupling the QCD axion to a new 
$U(1)'$ gauge group.  If the reheating of the universe produces magnetic 
monopoles under this $U(1)'$, the monopole density generates a mass 
for the axion~\cite{Fischler:1983sc}.  This is because topological 
terms like $F\tilde{F}$ become physical in the presence of magnetic 
monopoles due to the Witten effect~\cite{Witten:1979ey}.  Specifically, 
it gives a free energy density that depends on a background axion field 
value, thus creating an effective mass for the axion.  This mass is 
sufficient to damp isocurvature perturbations in the axion field.  After 
the perturbations have been damped, the monopole density can be efficiently 
eliminated by breaking the $U(1)'$ symmetry, resulting in confinement 
and subsequent annihilation of the monopoles.  The monopole density 
forces the axion field to relax into $\theta'$, a point on the potential 
chosen by $CP$ phases in the $U(1)'$ sector.  Since this phase need 
not be aligned with the QCD minimum at $\theta_{\rm QCD}$, the axion 
generally acquires a homogeneous cosmic abundance during the QCD phase 
transition, with suppressed isocurvature perturbations.  For large 
$f_a \gg 10^{12}~{\rm GeV}$, this misalignment needs to be small, 
$|\theta' - \theta_{\rm QCD}| \ll 1$, but this can be environmentally 
selected~\cite{Linde:1987bx}.  We show that there is sufficient time 
for the damping of axion isocurvature fluctuations so that axion 
dark matter with a unification scale $f_a$ is consistent with high 
scale inflation giving an observable size of the gravitational wave 
polarization signal.

The organization of this paper is as follows.  In Section~\ref{sec:damping}, 
we review the required amount of damping of axion isocurvature fluctuations 
consistent with current observations.  In Section~\ref{sec:basic} we 
introduce our basic mechanism, and in Section~\ref{sec:minimal} we 
present a minimal model realizing it.  We show that the model can 
consistently accommodate unification scale axion dark matter with high 
scale (unification scale) inflation.  In Section~\ref{sec:mono-annh}, we 
discuss monopole annihilations due to $U(1)'$ breaking in detail, showing 
that they efficiently eliminate monopoles.  In Section~\ref{sec:natural}, 
we discuss extensions/modifications of the minimal model in which the 
issue of radiative stability of the $U(1)'$ sector existing in the minimal 
model does not arise.  We conclude in Section~\ref{sec:concl}.

\section{Required Damping of Isocurvature Perturbations}
\label{sec:damping}

Inflation generally induces quantum fluctuations of order $H_{\rm inf}/2\pi$ 
for any massless field, where $H_{\rm inf}$ is the Hubble parameter during 
inflation.  This implies that if $U(1)_{\rm PQ}$ is broken before or during 
inflation, then the angle $\theta = a/f_a$ of the axion field $a$ has 
fluctuations
\begin{equation}
  \delta\theta(T_{\rm R}) \approx \frac{H_{\rm inf}}{2\pi f_a},
\label{eq:d-theta_init}
\end{equation}
at temperature $T_{\rm R}$, when the radiation dominated era starts 
due to reheating.%
\footnote{In this paper we adopt the instant reheating approximation 
 for simplicity, so that the universe is radiation dominated right 
 after inflation.  An extension of our analysis to more general cases 
 (including a matter dominated era before reheating) is straightforward.}
Since the axion potential is flat during inflation, these fluctuations 
are converted to isocurvature density perturbations upon the generation 
of the axion mass.

There is a tight constraint on the amount of allowed isocurvature 
perturbations from the Planck data~\cite{Planck:2013jfk}, which can 
be written as (see, e.g.,~\cite{Choi:2015zra})
\begin{equation}
  \frac{\Omega_a}{\Omega_{\rm DM}} 
    \frac{\delta\theta(T_{\rm QCD})}{\theta_{\rm mis}} 
  \lesssim 4.8 \times 10^{-6},
\label{eq:iso-bound}
\end{equation}
where $\theta_{\rm mis}$ is the average axion misalignment angle, 
while $\delta\theta(T_{\rm QCD})$ is the angle fluctuation of the 
axion field at temperature $T_{\rm QCD} \sim 1~{\rm GeV}$.  Here, 
$\Omega_a$ and $\Omega_{\rm DM} \simeq 0.24$ represent the axion 
and total dark matter abundances, respectively, and we assume 
$\theta_{\rm mis} > \delta\theta(T)$ throughout.%
\footnote{This condition requires $H_{\rm inf} \lesssim 2 \times 
 10^{14}~{\rm GeV} \sqrt{\Omega_a/\Omega_{\rm DM}} (f_a/10^{16}~{\rm 
 GeV})^{0.4}$; for comparison, see Eq.~(\ref{eq:H_inf-bound}) and 
 an estimate below it for unification scale inflation.}
Using the expression for the axion relic density
\begin{equation}
  \frac{ \Omega_a}{\Omega_{DM} }
  \approx 1.0 \times 10^5\, \theta_{\rm mis}^2 
    \left( \frac{f_a}{10^{16}~{\rm GeV}} \right )^{1.19},
\label{eq:a-relic}
\end{equation}
(which requires $\theta_{\rm mis} \lesssim 0.003$ for $f_a \simeq 
10^{16}~{\rm GeV}$, possibly realized through environmental selection 
effects~\cite{Linde:1987bx}), we may rewrite Eq.~(\ref{eq:iso-bound}) as
\begin{equation}
  \delta\theta(T_{\rm QCD}) 
  \lesssim 1.5 \times 10^{-8} \sqrt{\frac{\Omega_{\rm DM}}{\Omega_a}} 
    \left( \frac{10^{16}~{\rm GeV}}{f_a} \right)^{0.6}.
\label{eq:d_theta-bound}
\end{equation}
Assuming the standard cosmological history after inflation, 
$\delta\theta(T_{\rm QCD}) \approx \delta\theta(T_{\rm R})$, so that 
we find
\begin{equation}
  H_{\rm inf} \lesssim 9.4 \times 10^8~{\rm GeV} 
    \sqrt{\frac{\Omega_{\rm DM}}{\Omega_a}} 
    \left( \frac{f_a}{10^{16}~{\rm GeV}} \right)^{0.4}.
\label{eq:H_inf-bound}
\end{equation}
This severely constrains inflationary models in the presence of a 
unification scale axion~\cite{Fox:2004kb}.  In particular, unification 
scale axion dark matter---$\Omega_a = \Omega_{\rm DM}$ and $f_a 
\sim 10^{16}~{\rm GeV}$---is inconsistent with unification scale 
inflation---$E_{\rm inf} \equiv V_{\rm inf}^{1/4} \sim 10^{16}~{\rm GeV}$, 
which leads to $H_{\rm inf} = E_{\rm inf}^2/\sqrt{3} \bar{M}_{\rm Pl} 
\sim 10^{13}~{\rm GeV}$, where $\bar{M}_{\rm Pl} \simeq 2.4 \times 
10^{18}~{\rm GeV}$ is the reduced Planck scale.

Below, we discuss a scenario in which axion isocurvature fluctuations 
are damped due to dynamics after inflation.  Defining the (inverse) 
damping factor $\Delta$ by
\begin{equation}
  \Delta = \frac{\delta\theta(T_{\rm QCD})}{\delta\theta(T_{\rm R})},
\label{eq:epsilon}
\end{equation}
Eq.~(\ref{eq:d_theta-bound}) yields
\begin{equation}
  \Delta \lesssim 1 \times 10^{-4} 
    \sqrt{\frac{\Omega_{\rm DM}}{\Omega_a}} 
    \left( \frac{f_a}{10^{16}~{\rm GeV}} \right)^{0.4} 
    \left( \frac{10^{13}~{\rm GeV}}{H_{\rm inf}} \right).
\label{eq:epsilon-bound}
\end{equation}
Here, we have normalized $f_a$ and $H_{\rm inf}$ by the values corresponding 
to unification scale axion and inflation, respectively.  This gives the 
required amount of damping.

\section{Basic Mechanism}
\label{sec:basic}

Our basic idea of suppressing axion isocurvature fluctuations is that the 
axion mass obtains extra contributions beyond that from QCD in the early 
universe so that it is larger than the Hubble parameter in some period. 
In this period, axion isocurvature perturbations are reduced because of 
the damped oscillations of the axion field, giving $\Delta < 1$.

We do this by introducing a coupling of the axion to a hidden $U(1)'$ 
gauge group
\begin{equation}
  {\cal L} \sim \frac{1}{f_a}\, a F^{\prime\mu\nu} \tilde{F}^\prime_{\mu\nu}.
\label{eq:aFF}
\end{equation}
We assume that at some temperature $T_M$ after inflation ($T_M \lesssim 
T_{\rm R} \approx E_{\rm inf}$), monopoles of $U(1)'$ are created.  This 
can happen, for example, if a hidden sector $SU(2)'$ gauge group is broken 
to $U(1)'$ at that scale.%
\footnote{If the creation of monopoles is associated with $G \rightarrow 
 G' \times U(1)'$ symmetry breaking in the hidden sector, where $G$ and 
 $G'$ are non-Abelian gauge groups, then we would need to have two axion 
 fields in the ultraviolet so that the QCD axion remains after $G'$ 
 gives a large mass to one linear combination of the two axion fields.}
In the presence of magnetic monopoles, the coupling in Eq.~(\ref{eq:aFF}) 
induces an effective mass for the axion~\cite{Fischler:1983sc}:
\begin{equation}
  m_a^2(T) = \gamma \frac{n_M(T)}{f_a},
\label{eq:ma2}
\end{equation}
where $\gamma$ is determined by the structure of the $U(1)'$ sector, such 
as the gauge coupling and matter content.  ($\gamma$ may in general depend 
on temperature, although it is not the case in the explicit model considered 
below.)  $n_M(T)$ is the number density of the monopoles; assuming the 
abundance determined by the Kibble-Zurek mechanism~\cite{Kibble:1976sj}, 
we find
\begin{equation}
  n_M(T) \approx \alpha \left( \frac{T}{T_M} \right)^3 H(T_M)^3,
\label{eq:n_M}
\end{equation}
where $H(T)$ is the Hubble parameter at temperature $T$, and $\alpha 
\gtrsim 1$.%
\footnote{Note that $\alpha$ can be much larger than $O(1)$, depending on 
 the dynamics of the phase transition; see e.g.~\cite{Murayama:2009nj}. 
 In this case, monopole-antimonopole annihilations at $T \sim T_M$ may 
 become important; see Section~\ref{subsec:SUSY} for such a scenario.}
The contribution of Eq.~(\ref{eq:ma2}) makes the axion mass effect 
dominates over the Hubble friction
\begin{equation}
  m_a(T) \gtrsim 3 H(T),
\label{eq:ma_H-cond}
\end{equation}
below some temperature $T_{\rm i}$ ($\leq T_M$), so that the axion field 
is subject to damped oscillations for $T \lesssim T_{\rm i}$.

We assume that $U(1)'$ is spontaneously broken at some temperature 
$T_{\rm f}$ ($\ll T_{\rm i}$), so that monopoles quickly disappear.%
\footnote{An alternative possibility will be discussed in 
 Section~\ref{subsec:dyon}.}
Axion isocurvature fluctuations are then damped efficiently between 
temperatures $T_{\rm i}$ and $T_{\rm f}$.  Suppose
\begin{equation}
  m_a^2(T) \propto T^n,
\label{eq:ma2-Tn}
\end{equation}
($n=3$ for a constant $\gamma$).  Since the axion ``number density'' 
$m_a(T) \delta\theta(T)^2$ scales as $T^3$ while Eq.~(\ref{eq:ma_H-cond}) 
is satisfied, we find
\begin{equation}
  \delta\theta(T) \propto T^p,
\qquad
  p \approx \frac{6-n}{4},
\label{eq:deltheta-Tp}
\end{equation}
in this period.  The final damping factor is thus
\begin{equation}
  \Delta \approx \left( \frac{T_{\rm f}}{T_{\rm i}} \right)^{\frac{6-n}{4}},
\label{eq:epsilon-T}
\end{equation}
which can be compared with the required amount of damping from observations, 
Eq.~(\ref{eq:epsilon-bound}).

Note that the average axion field $\langle \theta \rangle = 
\langle a \rangle/f_a$ after the operation of this damping mechanism 
is determined by the structure of the hidden sector (the original hidden 
sector $\bar{\theta}$ parameter), which in general differs from the 
minimum of the late-time axion potential, $\theta_{\rm QCD}$.  A 
homogeneous displacement of the axion field from $\theta_{\rm QCD}$, 
determining the late-time axion dark matter abundance, is not controlled 
by the present mechanism, unless we make an extra assumption.  For 
$f_a \gg 10^{12}~{\rm GeV}$, the value of this displacement must be 
small, but it can be environmentally selected to be consistent with 
$\Omega_a \leq \Omega_{\rm DM}$~\cite{Linde:1987bx}.

\section{Minimal Model}
\label{sec:minimal}

We now consider the minimal model in which the $U(1)'$ sector below 
$T_M$ contains only a charged scalar field $\varphi$, which breaks 
$U(1)'$ at scale $T_{\rm f}$ ($\ll T_{\rm M}$).  In this case, 
the factor $\gamma$ in the expression for the induced axion mass, 
Eq.~(\ref{eq:ma2}), is
\begin{equation}
  \gamma \approx \tilde{\gamma}\, \frac{T_M}{f_a},
\label{eq:gamma-min}
\end{equation}
where we have used $T_M \lesssim f_a$, and $\tilde{\gamma} \approx O(1)$ 
assuming that the $U(1)'$ gauge coupling is of order unity.%
\footnote{It is important here that the $U(1)'$ sector does not contain 
 a light fermion charged under $U(1)'$.  If it did, virtual fermions 
 would partially screen the charge surrounding a monopole, allowing 
 it to spread over a distance or order $m_f^{-1}$.  Here, $m_f$ is the 
 fermion mass.  This would suppress the induced mass of the axion so 
 that $\gamma \approx m_f/f_a$~\cite{Fischler:1983sc}.  This will be 
 relevant for models in Section~\ref{subsec:SUSY}. \label{ft:ferm}}
The axion mass just after the monopole production is then given by
\begin{equation}
  \frac{m_a(T_M)}{3 H(T_M)} \simeq 0.2 \sqrt{\alpha \tilde{\gamma}}\, 
    g_{*M}^{\frac{1}{4}} \sqrt{\frac{T_M^3}{f_a^2 \bar{M}_{\rm Pl}}},
\label{eq:ma-H}
\end{equation}
where we have used $H(T_M) = \rho(T_M)^{1/2}/\sqrt{3} \bar{M}_{\rm Pl}$ 
and $\rho(T_M) = (\pi^2/30) g_{*M} T_M^4$ with $g_{*M}$ being the effective 
number of relativistic degrees of freedom at temperature $T_M$.  Assuming 
that $T_M$ is not much smaller than the unification scale, this number 
is roughly of order unity (and at least not too much smaller than of 
order unity).  The axion field thus starts having damped oscillations at
$T \sim T_i$, within a few orders of magnitude from $T_M$.  Specifically
\begin{equation}
  T_i \simeq 1 \times 10^{11}~{\rm GeV} \, \alpha \tilde{\gamma} 
    \sqrt{\frac{g_{*M}}{100}} \left( \frac{10^{16}~{\rm GeV}}{f_a} \right)^2 
    \left( \frac{T_M}{3 \times 10^{15}~{\rm GeV}} \right)^4.
\label{eq:Ti-min}
\end{equation}
Note that if $T_i$ in this expression exceeds $T_M$, e.g.\ because of 
$\alpha \gg 1$, then $T_{\rm i}$ must be set to $T_M$.

At temperatures below $T_{\rm i}$, axion isocurvature fluctuations 
are damped.  Since Eq.~(\ref{eq:gamma-min}) implies $n = 3$, so that 
$p \approx 3/4$ (see Eq.~(\ref{eq:deltheta-Tp})), 
\begin{equation}
  \frac{\delta\theta(T)}{\delta\theta(T_{\rm i})} 
  \approx \left( \frac{T}{T_{\rm i}} \right)^{\frac{3}{4}}.
\label{eq:damp-min}
\end{equation}
Therefore, to avoid the observational constraint of 
Eq.~(\ref{eq:epsilon-bound}), we need
\begin{equation}
  T_{\rm f} \lesssim 2 \times 10^5~{\rm GeV}\, 
    \alpha \tilde{\gamma} \sqrt{\frac{g_{*M}}{100}} 
    \left( \frac{\Omega_{\rm DM}}{\Omega_a} \right)^{\frac{2}{3}} 
    \left( \frac{T_M/E_{\rm inf}}{0.3} \right)^4 
    \left( \frac{10^{16}~{\rm GeV}}{f_a} \right)^{1.5} 
    \left( \frac{E_{\rm inf}}{10^{16}~{\rm GeV}} \right)^{\frac{4}{3}},
\label{eq:Tf-min}
\end{equation}
where we have used $H_{\rm inf} \approx E_{\rm inf}^2/\sqrt{3} 
\bar{M}_{\rm Pl}$.  We here generate the required value of $T_{\rm f}$ 
simply by the Brout-Englert-Higgs mechanism associated with $\varphi$:
\begin{equation}
  V_{\rm hid} = \lambda' \left( |\varphi|^2 - v^{\prime 2} \right)^2,
\label{eq:V_hidden}
\end{equation}
with $v' \approx T_{\rm f}$.  We find that unification scale axion 
dark matter with unification scale inflation can be made consistent 
by our mechanism.

Incidentally, ignoring $U(1)'$ breaking, we find that monopoles dominate 
the energy density of the universe at temperature
\begin{equation}
  T_* \simeq 6 \times 10^6~{\rm GeV}\, \alpha \sqrt{\frac{g_{*M}}{100}} 
    \left( \frac{T_M}{3 \times 10^{15}~{\rm GeV}} \right)^3 
    \left( \frac{m_M}{3 \times 10^{15}~{\rm GeV}} \right),
\label{eq:T_*}
\end{equation}
which is slightly below the upper bound in Eq.~(\ref{eq:Tf-min}) in 
the relevant parameter region.  Here, $m_M$ is the monopole mass.  This 
implies that the universe may be monopole dominated toward the end of 
the damped oscillation period, $T_{\rm f} \lesssim T \lesssim T_{\rm i}$.

\section{Monopole Annihilations}
\label{sec:mono-annh}

Here we discuss annihilations of monopoles after $U(1)'$ is spontaneously 
broken at some temperature $T_S$ ($\sim T_{\rm f}$).  After $U(1)'$ 
is spontaneously broken, monopoles and antimonopoles become connected 
by strings.  For monopole-antimonopole annihilations to occur, the 
string-monopole system must lose their energies, and there are several 
processes that can contribute to the energy loss.

We assume the existence of a renormalizable coupling between the $U(1)'$ 
and standard model sectors, e.g.\ a quartic coupling between the $U(1)'$ 
breaking and standard model Higgs fields or a kinetic mixing between 
$U(1)'$ and $U(1)$ hypercharge:
\begin{equation}
  {\cal L} \sim \epsilon\, \varphi^\dagger \varphi h^\dagger h,
\qquad
  \epsilon F'_{\mu\nu} F_Y^{\mu\nu}.
\label{eq:coup-sectors}
\end{equation}
We will find that monopoles quickly disappear, well within a Hubble time, 
unless the coupling $\epsilon$ is significantly suppressed.  Note that 
cosmic strings formed by $U(1)'$ breaking are harmless for $T_S \lesssim 
10^{15}~{\rm GeV}$~\cite{Seljak:2006bg}.

\subsection{Monopole friction}
\label{subsec:friction}

Suppose the correlation length of the $U(1)'$ breaking field, $\varphi$, 
is of order or larger than the average distance between monopoles at 
$T \sim T_S$:
\begin{equation}
  d(T_S) \sim n_M(T_S)^{-\frac{1}{3}} 
  \sim \frac{\bar{M}_{\rm Pl}}{\alpha^{1/3} T_S T_M}.
\label{eq:distance}
\end{equation}
In this case, strings will connect monopoles through the shortest 
possible path, and the energy of a monopole-antimonopole pair to be 
dissipated is
\begin{equation}
  E_0 \sim \eta\, d(T_S) 
  \sim \frac{\bar{M}_{\rm Pl} T_S}{\alpha^{1/3} T_M},
\label{eq:E_0}
\end{equation}
where we have estimated the string tension $\eta$ to be of order $T_S^2$.

If the monopoles scatter with a thermal bath of temperature $T_S$ through 
a coupling of strength $\epsilon$, as in Eq.~(\ref{eq:coup-sectors}), 
then the energy loss rate due to friction is~\cite{Vilenkin:2000jqa}:
\begin{equation}
  \dot{E} \sim - \epsilon^2 T_S^2 v^2,
\label{eq:E-dot}
\end{equation}
where $v$ is the velocity of the monopoles, which is given by
\begin{equation}
  v \sim \left\{ \begin{array}{ll}
  \left( \frac{T_S^2 d (T_S)}{m_M} \right)^{\frac{1}{2}} 
    \sim \left( \frac{T_S \bar{M}_{\rm Pl}}{\alpha^{1/3} T_M^2} 
    \right)^{\frac{1}{2}} 
  & \mbox{ for}\,\, T_S \ll \frac{\alpha^{1/3} T_M^2}{\bar{M}_{\rm Pl}},
\\
  1 & \mbox{ for}\,\, T_S \gtrsim 
    \frac{\alpha^{1/3} T_M^2}{\bar{M}_{\rm Pl}},
  \end{array} \right.
\label{eq:v_M}
\end{equation}
where the former and latter cases correspond to nonrelativistic and 
relativistic monopoles, respectively.  In each case, the annihilation 
timescale $\tau_{\rm ann} \sim |E_0/\dot{E}|$ is given by
\begin{equation}
  \tau_{\rm ann} \sim \left\{ \begin{array}{ll}
  \frac{T_M}{\epsilon^2 T_S^2} 
  & \mbox{ for}\,\, T_S \ll \frac{\alpha^{1/3} T_M^2}{\bar{M}_{\rm Pl}},
\\
  \frac{\bar{M}_{\rm Pl}}{\epsilon^2 \alpha^{1/3} T_S T_M} 
  & \mbox{ for}\,\, T_S \gtrsim \frac{\alpha^{1/3} T_M^2}{\bar{M}_{\rm Pl}}.
  \end{array} \right.
\label{eq:tau_ann}
\end{equation}
In both cases, this timescale is of order or shorter than the Hubble 
timescale, $t_S \sim \bar{M}_{\rm Pl}/T_S^2$, unless $\epsilon$ is much 
smaller than of order unity.

\subsection{Particle production from strings}
\label{subsec:prod-strings}

If the correlation length of $\varphi$ is much smaller than the average 
monopole distance at $T_S$, then we expect that a string connecting a 
monopole-antimonopole pair to have a significant number of kinks (from 
a Brownian formation), and particle production from the string contributes 
significantly to the dissipation.

Based on the analysis in Ref.~\cite{Vilenkin:2000jqa}, we estimate that 
the power for a string of thickness $\delta$ and length $L$ to radiate 
standard model particles is
\begin{equation}
  P \sim \frac{\epsilon^2}{\delta\, \xi(T_S)},
\label{eq:power}
\end{equation}
per a portion of a string of length $\xi(T_S)$, where $\xi(T_S)$ 
($\ll d(T_S)$) is the correlation length of $\varphi$.%
\footnote{The process of energy dissipation may be much faster, 
 $P \sim \epsilon^2 \eta (\delta/\xi(T_S))^{1/3}$, if cusps form 
 efficiently~\cite{Brandenberger:1986vj}.  Here we adopt a conservative 
 estimate of Eq.~(\ref{eq:power}), which is sufficient to eliminate 
 the monopoles quickly.}
In the case of Brownian strings, the average string length is given by
\begin{equation}
  L \sim \frac{d(T_S)^2}{\xi(T_S)},
\label{eq:av-L}
\end{equation}
so that the total energy of the string-monopole system to be dissipated 
and the emission power from it are
\begin{align}
  & E_0 \sim \eta L \sim \frac{T_S^2\, d(T_S)^2}{\xi(T_S)},
\label{eq:E_0-2}\\
  & \dot{E} \sim P \frac{L}{\xi(T_S)} 
  \sim \frac{\epsilon^2 T_S\, d(T_S)^2}{\xi(T_S)^3},
\label{E_dot-2}
\end{align}
where we have used $\eta \sim T_S^2$ and $\delta \sim 1/T_S$.  The 
monopole-antimonopole annihilation timescale is thus
\begin{equation}
  \tau_{\rm ann} = \frac{E_0}{\dot{E}} 
  \sim \frac{1}{\epsilon^2} T_S\, \xi(T_S)^2 
  \ll \frac{1}{\epsilon^2} T_S\, d(T_S)^2 
  \sim \frac{\bar{M}_{\rm Pl}^2}{\epsilon^2 \alpha^{2/3} T_S T_M^2}.
\label{eq:tau_ann-2}
\end{equation}
Again, this is of order or shorter than the Hubble timescale, $t_S 
\sim \bar{M}_{\rm Pl}/T_S^2$, unless $\epsilon$ is much smaller than 
order unity.%
\footnote{In the analysis in this subsection, we have ignored the effect 
 of the increase of the relevant correlation length due to interactions 
 of the strings with the thermal bath, which may become important for 
 $T_S \lesssim T_M^2/\bar{M}_{\rm Pl}$.  In this case, however, the 
 analysis in the previous subsection applies, which also says that 
 monopoles quickly disappear after $U(1)'$ symmetry breaking.}

\section{Technical Naturalness of {\boldmath $U(1)'$}}
\label{sec:natural}

In Section~\ref{sec:minimal}, we have presented the minimal model in which 
$U(1)'$ breaking is achieved by a scalar field $\varphi$ with the potential 
Eq.~(\ref{eq:V_hidden}).  As it stands, the scale appearing in this 
potential, $v'$, is not radiatively stable.  The radiative stability 
of this scale is qualitatively and quantitatively different from the 
problem of protecting the QCD axion from quantum corrections.  The 
$U(1)'$-breaking field $\varphi$ is a scalar much like the standard 
model Higgs field whose mass needs to be protected at scales above 
$T_S$, unlike the QCD axion whose mass needs to be protected to the 
level of $\sim 10^{-5} m_a$.  Existing ideas to address the hierarchy 
problem may thus be leveraged to solve this issue.  In this section, 
we discuss extensions/modifications of the minimal model in which the 
issue of radiative stability does not arise.

\subsection{Supersymmetric {\boldmath $U(1)'$} sector}
\label{subsec:SUSY}

One way to construct a technically natural model is to make the $U(1)'$ 
sector supersymmetric.  This requires promoting the $U(1)'$-breaking 
field $\varphi$ to chiral superfields $\Phi(+1)$ and $\bar{\Phi}(-1)$. 
The complication arises because the induced axion mass is suppressed 
in the presence of light fermions charged under $U(1)'$, as mentioned 
in footnote~\ref{ft:ferm}.  To obtain a significant contribution to 
the axion mass, we need to have a supersymmetric mass for $\Phi$ and 
$\bar{\Phi}$:
\begin{equation}
  W = M_\Phi \Phi \bar{\Phi}.
\label{eq:M_Phi}
\end{equation}
The breaking of $U(1)'$ is then caused by supersymmetry-breaking squared 
masses for $\Phi$ and $\bar{\Phi}$ of order $\tilde{m}^2 \sim T_S^2$. 
To maximize the axion mass, we also take $M_\Phi \sim T_S$.%
\footnote{The coincidence of the scales $\tilde{m}$ and $M_\Phi$ is 
 analogous to the $\mu$ problem in the minimal supersymmetric standard 
 model, which can be addressed, e.g., as in Ref.~\cite{Giudice:1988yz}.}
The coupling between the $U(1)'$ and the standard model sectors needed 
for monopole annihilations can be taken as a kinetic mixing between $U(1)'$ 
and $U(1)$ hypercharge: ${\cal L} \sim \epsilon\, [{\cal W}^{\prime\alpha} 
{\cal W}_{Y \alpha}]_{\theta^2}$ (see Section~\ref{sec:mono-annh}). 
This implies that the standard model is also supersymmetric above the 
scale $\sim (4\pi/\epsilon) \tilde{m}$.

With this setup, the induced axion mass is given by Eq.~(\ref{eq:ma2}) 
with
\begin{equation}
  \gamma \approx \frac{M_\Phi}{f_a} \sim \frac{T_S}{f_a}.
\label{eq:gamma-SUSY}
\end{equation}
Plugging this into Eq.~(\ref{eq:Tf-min}) with $T_S \sim T_{\rm f}$, we 
find that $\alpha$ must be much larger than $1$ for the model to work. 
We thus suppose that the dynamics of the phase transition producing 
monopoles is such that $\alpha \gg 1$.  The largest possible abundance 
of monopoles obtained in this case is determined by the freezeout 
abundance (instead of Eq.~(\ref{eq:n_M})), which is given 
by~\cite{Preskill:1979zi}
\begin{equation}
  n_M(T) \approx \left( \frac{T}{T_M} \right)^3 
    \frac{\sqrt{g_{*M}} T_M^4}{\bar{M}_{\rm Pl}},
\label{eq:n_M-freezeout}
\end{equation}
where we have assumed an $O(1)$ $U(1)'$ gauge coupling.  The axion mass 
at $T \sim T_M$ is then
\begin{equation}
  \frac{m_a(T_M)}{3 H(T_M)} 
  \sim \frac{\sqrt{T_S \bar{M}_{\rm Pl}}}{g_{*M}^{1/4} f_a},
\label{eq:ma-H-SUSY}
\end{equation}
so that the axion field starts damped oscillations at
\begin{equation}
  T_{\rm i} \sim \frac{T_S T_M \bar{M}_{\rm Pl}}{\sqrt{g_{*M}} f_a^2}.
\label{eq:Ti-SUSY}
\end{equation}
This gives the damping factor of
\begin{equation}
  \Delta \approx \left( \frac{T_{\rm f}}{T_{\rm i}} \right)^{\frac{3}{4}} 
  \sim \left( \frac{f_a^2}{T_M \bar{M}_{\rm Pl}} \right)^{\frac{3}{4}}.
\label{eq:Delta-SUSY}
\end{equation}
We find that the mechanism is not as strong as in the minimal model, but 
it can still save the scenario with $f_a$, $T_M$, $E_{\rm inf}$ as large 
as $\sim 10^{15}~{\rm GeV}$.

\subsection{Possibility of unbroken {\boldmath $U(1)'$}}
\label{subsec:dyon}

We finally mention an alternative (and very different) possibility that 
$U(1)'$ monopoles may be efficiently eliminated without breaking $U(1)'$. 
This may happen if the monopole under consideration is in fact a dyon 
that also carries a charge under a hidden non-Abelian gauge group $G'$ 
(to which the axion field does not couple).  In this case, if $G'$ 
confines at a scale $\Lambda'$, then dyons can be subjected to extra 
strong annihilation processes.

Suppose the $G'$ sector contains light particles that are electrically 
charged under $G'$.  When $G'$ confines at $T \sim \Lambda'$, dyons 
pick up these light particles, becoming $G'$ hadrons.  At this point, 
the dyon-antidyon annihilation cross section is expected to become 
large $\sim 1/\Lambda^{\prime 2}$, as in the analogous situation 
for a heavy stable colored particle~\cite{Kang:2006yd}.  This will 
efficiently eliminate dyons if the confinement scale is sufficiently 
low $\Lambda' \lesssim 100~{\rm TeV}$, giving $T_{\rm f} \sim \Lambda'$. 
Since this scenario does not require breaking of the $U(1)'$ symmetry, 
the $U(1)'$ sector need not have a light charged scalar or fermion, 
which would, respectively, lead to the issue of radiative stability 
and axion mass suppression.  Further studies of this possibility, 
including a detailed analysis of whether dyon annihilation is indeed 
strong enough, are warranted.

\section{Conclusions}
\label{sec:concl}

In this paper we have presented a mechanism that suppresses axion 
isocurvature fluctuations due to the dynamics of a hidden $U(1)'$ 
sector coupled to the axion field.  In particular, this sector produces 
$U(1)'$ monopoles at $T \sim T_M$, which disappear at $T \sim T_{\rm f}$ 
($\ll T_M$).  For temperatures between $T_{\rm i}$ ($\gg T_{\rm f}$) 
and $T_{\rm f}$, the effective axion mass induced by the monopoles makes 
the axion heavier than the Hubble parameter, so that the isocurvature 
fluctuations are damped.  Since the average value of the axion field 
after the damping is not necessarily at the minimum of the zero-temperature 
potential determined by QCD, homogeneous coherent oscillations 
after the QCD phase transition may still produce axion dark 
matter~\cite{Linde:1987bx}.

We have presented a minimal model in which this mechanism successfully 
operates.  This model accommodates a large enough time interval in which 
the axion isocurvature fluctuations are damped, so that axion dark matter 
with a unification scale decay constant can be consistent with unification 
scale inflation.  We have also discussed extensions/modifications of the 
minimal model in which the issue of radiative stability does not arise.

Since the axion provides a leading solution to the strong $CP$ problem, 
it is important to fully study its consistency.  If a future CMB 
experiment discovers inflationary gravitational wave signals, it 
would exclude naive axion models with the Peccei-Quinn symmetry broken 
before the end of inflation.  Our mechanism makes the QCD axion alive 
even in such a case, without requiring the Peccei-Quinn symmetry 
breaking scale to be below the inflationary scale.  This is particularly 
important for a string axion, which has a virtue that explicit 
breaking of the Peccei-Quinn symmetry (which needs to be extremely 
small to solve the strong $CP$ problem~\cite{Kamionkowski:1992mf}) 
is generated only at a nonperturbative level~\cite{Svrcek:2006yi}. 
Our mechanism allows for a string axion to be a consistent solution 
to the strong $CP$ problem even if inflationary gravitational wave 
signals are discovered, and it would also keep open the possibility 
that axion dark matter may be discovered by high precision experiments 
such as those proposed in Ref.~\cite{Graham:2011qk}.

\vspace{2mm}
\begin{flushleft}
{\bf Note added:}

While completing this paper, we received Ref.~\cite{KTY} which discusses 
a similar idea.
\end{flushleft}

\section*{Acknowledgments}

We would like to thank Peter Graham, Nemanja Kaloper, David E. Kaplan, 
Jeremy Mardon, and Sean Weinberg for discussions.  This work was 
supported in part by the Director, Office of Science, Office of High 
Energy and Nuclear Physics, of the U.S.\ Department of Energy (DOE) under 
Contract DE-AC02-05CH11231.  Y.N. was supported in part by the National 
Science Foundation (NSF) under grant PHY-1521446 and MEXT KAKENHI 
Grant Number 15H05895.  S.R. was supported in part by the NSF under 
grants PHY-1417295 and PHY-1507160, the Simons Foundation Award 378243, 
and the Heising Simons Foundation.  The work of F.S. was supported in 
part by the DOE National Nuclear Security Administration Stewardship 
Science Graduate Fellowship.

\end{document}